\documentclass[
aps,prd,
12pt,
nopreprintnumbers,
showpacs,
eqsecnum,
nofootinbib
]{revtex4-1}

\usepackage{graphicx}

\begin{document}

\title{
Critical Higher Order Gravities in Higher Dimensions}
\author{Nahomi Kan}\email[]{kan@gifu-nct.ac.jp}
\affiliation{
Gifu National College of Technology,
Motosu-shi, Gifu 501-0495, Japan
}
\author{Koichiro Kobayashi}\email[]{m004wa@yamaguchi-u.ac.jp}
\author{Kiyoshi Shiraishi}\email[]{shiraish@yamaguchi-u.ac.jp}
\affiliation{
Yamaguchi University,
Yamaguchi-shi, Yamaguchi 753--8512, Japan}
\date{\today}

\begin{abstract}
We show that the higher order gravity model proposed by Meissner and
Olechowski has a graviton mode, a massive spin-two excitation and no
scalar mode in a maximally symmetric spacetime; therefore, by choosing
the coefficients,  we can construct a Lagrangian for
`critical gravity' from higher order terms of curvatures in higher
dimensions. We also give a comment on construction of the theory with
multi-criticality in higher order gravities. 
\end{abstract}



\maketitle

\section{Introduction}

Effective theory of gravity may contain higher-derivative terms in its
Lagrangian density, in addition to the Einstein-Hilbert term.
The correction of the gravitational Lagrangian is motivated from studies
on quantum theory of gravitation and string theory.
Some special forms of higher order terms attract much attention nowadays.

In $D$-dimensional Lovelock gravity, whose the Lagrangian is written in
the dimensionally continued Euler densities
\cite{Lanczos,Lovelock,Zumino,Zwi,BD,JTWheeler,Aragone,MH,PK},
the second derivative of the metric tensor
disappears in the action. 
The excitation mode in the Lovelock gravity is only a 
(traceless, transverse) graviton mode in the $D$-dimensional Einstein
background metric, as has recently been shown by Camanho, Edelstein
and Paulos \cite{CEP}, Camanho and Edelstein \cite{CE}, and 
\c{S}i\c{s}man,
G\"ull\"u and Tekin \cite{SGT2}.

On the other hand, Meissner and Olechowski have found the
particular higher order invariants of the Ricci tensors and
scalar curvatures which effectively coincide with the dimensionally
continued Euler form if the spacetime is conformally flat
\cite{MO,MO2}.  Therefore, gravity theory governed by the Lagrangian
constructed from a linear combination of such invariants has no scalar
modes as the Einstein gravity.

In new massive gravity in three dimensions \cite{NMG,NMG2} and 
critical gravity theory recently proposed in
\cite{SGT,LP,DLLP,LPP,SGT1,Ohta}, scalar modes are absent from the
particle content of the theory. 
In the present paper, we generalize the structure of the
Lagrangian of the critical gravity to models with higher order terms in 
curvature tensors in $D$ dimensions. We show that such an extension can
be attained by use of the curvature polynomials introduced by Meissner
and Olechowski.

The outline of this paper is as follows.
In \S 2, we construct the higher order term in curvatures by
generalizing that in the Lagrangian of the critical gravity. In \S 3, The
metric fluctuation on a maximally symmetric spacetime is analyzed in the
special class of higher order gravity.  We propose  an extension of the
critical gravity to more higher-derivative theory in
\S4. 
The final
section is devoted to conclusion, where further prospect is also
addressed.

\section{Lovelock and Meissner-Olechowski gravity}
We shall begin with introducing the
dimensionally continued Euler density:
\begin{equation}
{L^{(n)}_L}=2^{-n}
\delta^{\sigma_1\tau_1\cdots\sigma_n\tau_n}_{\lambda_1\rho_1\cdots\lambda_n\rho_n}
R^{\lambda_1\rho_1}{}_{\sigma_1\tau_1}\cdots
R^{\lambda_n\rho_n}{}_{\sigma_n\tau_n}\,,
\end{equation}
where $R^{\mu\nu}{}_{\alpha\beta}$ is the Riemann tensor and the
generalized Kronecker delta is defined as
\begin{equation}
\delta^{\mu_1\mu_2\cdots\mu_p}_{\nu_1\nu_2\cdots\nu_p}
\equiv\left|\begin{array}{cccc}
\delta^{\mu_1}_{\nu_1} & \delta^{\mu_1}_{\nu_2} & \cdots &
\delta^{\mu_1}_{\nu_p} \\
\delta^{\mu_2}_{\nu_1} & \delta^{\mu_2}_{\nu_2} & \cdots &
\delta^{\mu_2}_{\nu_p} \\
\vdots & \vdots & \ddots & \vdots \\
\delta^{\mu_p}_{\nu_1} & \delta^{\mu_p}_{\nu_2} & \cdots &
\delta^{\mu_p}_{\nu_p} 
\end{array}\right|\,.
\end{equation}
One should note that
\begin{equation}
\delta^{\mu\mu_1\mu_2\cdots\mu_p}_{\mu\nu_1\nu_2\cdots\nu_p}
=(D-p)\delta^{\mu_1\mu_2\cdots\mu_p}_{\nu_1\nu_2\cdots\nu_p}\,,
\end{equation}
where $D$ denotes the dimension of the spacetime; namely, all the indices
take $\mu, \nu,\dots=0, 1, \dots , D-1$. The
dimensionally continued Euler density $L^{(n)}_L$ consists of $n$-th order
in the curvature tensors. For example, for $n=1$, we find the
Einstein-Hilbert term
\begin{equation}
{L^{(1)}_L}=R\,,
\end{equation}
where $R$ is the scalar curvature, and for $n=2$, we find the Gauss-Bonnet
term
\begin{equation}
{L^{(2)}_L}=R_{\mu\nu\rho\sigma}R^{\mu\nu\rho\sigma}-4
R_{\mu\nu}R^{\mu\nu}+R^2\,,
\end{equation}
where $R_{\mu\nu}$ is the Ricci tensor.
The Lovelock gravity is described by the Lagrangian density of a linear
combination of the dimensionally continued Euler 
forms \cite{Lovelock,Zumino,Zwi,JTWheeler,Aragone,MH,PK}.

Next, we consider
the Schouten tensor (in our definition, which is different from the
original definition by a factor
$(D-2)^{-1}$,) which is known as
\begin{equation}
S^{\mu\nu}=R^{\mu\nu}-\frac{1}{2(D-1)}Rg^{\mu\nu}\,.
\end{equation}
The $n$-th order Meissner-Olechowski density is defined here by using the
Schouten tensor and the generalized Kronecker delta as \cite{MO,MO2,KKS}
\begin{equation}
L_{MO}^{(n)}=-\delta^{\mu_1\cdots\mu_n}_{\nu_1\cdots\nu_n}
{S}^{\nu_1}_{\mu_1}\cdots{S}^{\nu_n}_{\mu_n}
\,.
\end{equation}
In the present paper, we call the theory governed by the Lagrangian
density of a linear combination of this type as the
Meissner-Olechowski gravity.

Particularly, for $n=2$, we find the following quadratic combination in
curvatures:
\begin{equation}
L_{MO}^{(2)}=
\left[R^{\mu\nu}-\frac{1}{2(D-1)}R\,
g^{\mu\nu}\right]\left(R_{\mu\nu}-\frac{1}{2}R\,
g_{\mu\nu}\right)=R_{\mu\nu}R^{\mu\nu}-\frac{D}{4(D-1)}R^2
\,.
\end{equation}
This scalar invariant is adopted in the argument on critical gravities in
three and four dimensions \cite{NMG,NMG2,SGT,LP,DLLP,LPP,SGT1,Ohta}.
Thus, the Meissner-Olechowski gravity is a possible candidate for an
extension of the critical gravity.

\section{analysis on metric fluctuation}
We consider perturbative metric fluctuation around a static, curved
background in this section. One should find the linearized gravitational
field equation in order to identify the particle content in the theory. 
Assuming that the metric is divided into the background and fluctuation,
we write
\begin{equation}
g_{\mu\nu}=\bar{g}_{\mu\nu}+h_{\mu\nu}\,.
\end{equation}
The indices are raised and lowered by the background metric $\bar{g}$.
Then the trace of the fluctuation is expressed as
\begin{equation}
h\equiv\bar{g}^{\mu\nu}h_{\mu\nu}\,.
\end{equation}

The following expansions to the quadratic order in $h_{\mu\nu}$ are
generally known \cite{BC}:
\begin{equation}
\int
d^Dx\sqrt{-g}=\int
d^Dx\sqrt{-\bar{g}}\left[1+\frac{1}{2}h+\frac{1}{8}
(h^2-2h^{\mu\nu}h_{\mu\nu})+O(h^3)\right]\,,
\end{equation}
\begin{eqnarray}
& &\int
d^Dx\sqrt{-g}R=\int
d^Dx\sqrt{-\bar{g}}\left(\frac{1}{2}\bar{R}\bar{g}^{\mu\nu}-\bar{R}^{\mu\nu}
\right)h_{\mu\nu}\nonumber \\
& &+\int
d^Dx\sqrt{-\bar{g}}\left[\frac{1}{4}h^{\mu\nu}\bar{\nabla}^2h_{\mu\nu}
+\frac{1}{2}h\bar{\nabla}_\mu\bar{\nabla}_\nu h^{\mu\nu}
-\frac{1}{2}h_{\mu\nu}\bar{\nabla}^\mu\bar{\nabla}_\lambda
h^{\lambda\nu}-\frac{1}{4}h\bar{\nabla}^2h+\frac{1}{2}h^{\mu\nu}\bar{R}_\mu{}^\alpha{}_\nu{}^\beta
h_{\alpha\beta}\right.\nonumber \\
&
&\qquad\qquad\qquad\left.+\frac{1}{2}h^{\mu\nu}\bar{R}_\mu^\alpha
h_{\alpha\nu}-\frac{1}{2}h\bar{R}^{\mu\nu}h_{\mu\nu}-
\frac{1}{4}\bar{R}h^{\mu\nu}h_{\mu\nu}+\frac{1}{8}\bar{R}h^2+O(h^3)\right]\,,
\end{eqnarray}
where $\bar{R}_{\mu\nu\lambda\rho}$, $\bar{R}_{\mu\nu}$, and
$\bar{R}$ are the Riemann tensor, the Ricci tensor and the scalar
curvature derived only from the background metric
$\bar{g}_{\mu\nu}$.

Now, we consider a $D$-dimensional maximally symmetric spacetime as the
background. Then, the Riemann tensor can be written as
\begin{equation}
\bar{R}^{\mu\nu}{}_{\alpha\beta}=\Lambda
(\delta^{\mu}_{\alpha}\delta^{\nu}_{\beta}-
\delta^{\mu}_{\beta}\delta^{\nu}_{\alpha})
\,,
\label{bg}
\end{equation}
where $\Lambda$ is a constant.
This background geometry is the solution of the Einstein equation obtained
by varying the action
\begin{equation}
S_0(\Lambda)=\int d^Dx\,\sqrt{-g}\,[R-(D-1)(D-2)\Lambda]
\,.
\end{equation}

The Riemann tensor, the Ricci tensor and the scalar curvature are written
to the first order in
$h_{\mu\nu}$ as
\begin{equation}
R^{\mu\nu}{}_{\alpha\beta}=\bar{R}^{\mu\nu}{}_{\alpha\beta}+
{\cal R}^{\mu\nu}{}_{\alpha\beta}+O(h^2)\,,\quad
R^{\mu}_{\alpha}=\bar{R}^{\mu}_{\alpha}+
{\cal R}^{\mu}_{\alpha}+O(h^2)\,,\quad
R=\bar{R}+
{\cal R}+O(h^2)\,,
\end{equation}
(here and hereafter, the abbreviation such that $O(h)=O(h_{\mu\nu})$ is
used), where
\begin{eqnarray}
{\cal
R}^{\mu\nu}{}_{\alpha\beta}&=&-\frac{1}{2}\Lambda(\delta^\mu_\alpha
h^\nu_\beta -\delta^\mu_\beta h^\nu_\alpha-\delta^\nu_\alpha
h^\mu_\beta+\delta^\nu_\beta h^\mu_\alpha)\nonumber \\
& &+\frac{1}{2}(-\bar{\nabla}_\alpha\bar{\nabla}^\mu h^\nu_\beta+
\bar{\nabla}_\alpha\bar{\nabla}^\nu h^\mu_\beta
+\bar{\nabla}_\beta\bar{\nabla}^\mu h^\nu_\alpha-
\bar{\nabla}_\beta\bar{\nabla}^\nu h^\mu_\alpha)\,, \\
{\cal
R}^{\mu}_{\alpha}&=&
\Lambda(h^{\mu}_{\alpha}-\delta^{\mu}_{\alpha}h)
+\frac{1}{2}(-\bar{\nabla}^2 h^\mu_\alpha-
\bar{\nabla}_\alpha\bar{\nabla}^\mu h+
\bar{\nabla}^\mu\bar{\nabla}_\nu h^\nu_\alpha
+\bar{\nabla}_\alpha\bar{\nabla}_\nu h^{\nu\mu})\,,
\label{ricci}
\\
{\cal
R}&=&
-(D-1)\Lambda h
+\bar{\nabla}_\mu\bar{\nabla}_\nu h^{\mu\nu}-\bar{\nabla}^2
h\,.
\label{scalar}
\end{eqnarray}

The Schouten tensor introduced in the previous section is also expanded
 by the order of $h_{\mu\nu}$ around the background. We define
the fluctuated part of the Schouten tensor as
\begin{equation}
S_\Lambda{}^{\mu}_{\alpha}\equiv S^{\mu}_{\alpha}-
\bar{S}^{\mu}_{\alpha}=R^{\mu}_{\alpha}-\frac{1}{2(D-1)}R
\delta^{\mu}_{\alpha}
-\frac{D-2}{2}\Lambda
\delta^{\mu}_{\alpha}={\cal R}^{\mu}_{\alpha}-\frac{1}{2(D-1)}{\cal R}
\delta^{\mu}_{\alpha}+O(h^2)\,,
\end{equation}
where the barred quantities indicates that it is estimated in the
background metric, or, the quantity in $O(h^0)$.
Using this tensor instead of the `whole' Schouten tensor, we find that
the second order Meissner-Olechowski density becomes
\begin{eqnarray}
L_{MO}^{(2)}{}_\Lambda&\equiv
&-\delta_{\mu\nu}^{\alpha\beta}{S_\Lambda}{}^\mu_\alpha
{S_\Lambda}{}^\nu_\beta\nonumber \\
&=&\left[R^\mu_\alpha-\frac{1}{2(D-1)}R\delta^\mu_\alpha
-\frac{D-2}{2}\Lambda\delta^\mu_\alpha\right]
\left[R_\mu^\alpha-\frac{1}{2}R\delta_\mu^\alpha+
\frac{(D-1)(D-2)}{2}\Lambda\delta_\mu^\alpha\right]\nonumber \\
&=&L_{MO}^{(2)}+\frac{(D-2)^2}{2}\Lambda
R-\frac{D(D-1)(D-2)^2}{4}\Lambda^2\,.
\label{eq1}
\end{eqnarray}
On the other hand, since $S_\Lambda{}^\mu_\alpha$ is obviously of linear
and higher order in
$h_{\mu\nu}$, it is clear that
\begin{equation}
L_{MO}^{(2)}{}_\Lambda={\cal
R}^{\mu}_{\alpha}{\cal
R}^{\alpha}_{\mu}-\frac{D}{4(D-1)}{\cal R}^2+O(h^3)\,,
\end{equation}
i.e., ${\cal
R}^{\mu}_{\alpha}{\cal
R}^{\alpha}_{\mu}-\frac{D}{4(D-1)}{\cal R}^2$ is the quantity of $O(h^2)$.
The explicit calculation using (\ref{ricci}) and (\ref{scalar}) yields
\begin{eqnarray}
{\cal
R}^{\mu}_{\alpha}{\cal
R}^{\alpha}_{\mu}-\frac{D}{4(D-1)}{\cal R}^2
&\simeq&
\Lambda^2\left[h^{\mu}_{\alpha}h^{\alpha}_{\mu}+(D-2)h^2-
\frac{1}{4}D(D-1)h^2\right]-
\frac{D-3}{4}\Lambda h \bar{\nabla}^2h\nonumber
\\ & &
-\Lambda h^{\alpha}_{\mu}\bar{\nabla}^2 h^\mu_\alpha
-(D-1)\Lambda
h_\mu^\alpha\bar{\nabla}^\mu\bar{\nabla}_\nu
h^\nu_\alpha+\frac{D}{2}\Lambda
h\bar{\nabla}_\mu\bar{\nabla}_\nu h^{\mu\nu}\nonumber \\
& &+\frac{1}{4}\left[h_\mu^\alpha\bar{\nabla}^2\bar{\nabla}^2 
h^\mu_\alpha-(\bar{\nabla}_\mu\bar{\nabla}_\nu
h^{\mu\nu})^2-\frac{1}{D-1}
(\bar{\nabla}_\mu\bar{\nabla}_\nu
h^{\mu\nu}-\bar{\nabla}^2 h)^2\right.
\nonumber \\
& &\left.-
(
\bar{\nabla}^\mu\bar{\nabla}_\nu h^\nu_\alpha
-\bar{\nabla}_\alpha\bar{\nabla}_\nu h^{\nu\mu})^2\right]
\,,
\end{eqnarray}
where `$\simeq$' denotes the equality up to the total derivatives.

Now, let us suppose the action
\begin{equation}
S=S_0(\Lambda)+\alpha\int d^Dx\,\sqrt{-g}\,L_{MO}^{(2)}{}_\Lambda\,.
\label{L2}
\end{equation}
We can analyze the linearized equation of motion for $h_{\mu\nu}$
by taking the variation of the action up to $O(h^2)$.
Applying the following gauge choice 
\begin{equation}
\bar{\nabla}_\nu
h^{\nu\mu}-\bar{\nabla}^\mu h=0\,,
\end{equation}
one can find the trace of the equation of motion derived from the action
(\ref{L2}) leads to
\begin{equation}
h=0\,.
\end{equation}
Therefore we get the following on-shell equation of motion for the metric
fluctuation, which is transverse and traceless ($\bar{\nabla}_\mu
h^{\mu\nu}=h=0$), on the maximally symmetric spacetime:
\begin{equation}
\alpha\left(\bar{\nabla}^2-2\Lambda\right)
\left(\bar{\nabla}^2-2\Lambda+\frac{1}{\alpha}\right)h_{\mu\nu}=0\,.
\label{2m}
\end{equation}
Note that the action (\ref{L2}) can be rewritten as
\begin{eqnarray}
& &S_0(\Lambda)+\alpha\int
d^Dx\,\sqrt{-g}\,L_{MO}^{(2)}{}_\Lambda\nonumber
\\ &=&\int
d^Dx\,\sqrt{-g}\,\left[\left\{1+\frac{(D-2)^2}{2}\alpha\Lambda\right\}R-
(D-1)(D-2)\left\{1+\frac{D(D-2)}{4}\alpha\Lambda\right\}\Lambda\right.\nonumber
\\
& &\qquad\qquad\qquad\left.+\alpha L^{(2)}_{MO}\right]\,,
\end{eqnarray}
and this is just the action considered in the original literature of
critical gravity, after rescaling the Newton and cosmological constants.

Eq.~(\ref{2m}) indicates that there are two tensor
excitation modes in general cases
\cite{SGT,LP,DLLP,LPP,SGT1,Ohta}.
Furthermore, one can find that
the critical point exists at $\alpha\rightarrow\infty$, or equivalently,
when the action then becomes $\int
d^Dx\,\sqrt{-g}\,L_{MO}^{(2)}{}_\Lambda$. At this point, a single
graviton excitation appears but it exists a log mode \cite{AF,BHRT,JNZ}%
\footnote{The first work on the log mode, in topologically massive
gravity in three dimension, has been done by Grumiller and
Johansson \cite{GJ1}. For a recent review, see Ref.~\cite{GRRZ}.}.

The argument so far is not new one but a lightning review of critical
gravity. We turn to generalize the construction of critical gravity by
using higher order terms in curvatures.
We consider the following higher order scalar invariants:
\begin{eqnarray}
L_{MO}^{(n+2)}{}_\Lambda
&\equiv&-\delta^{\mu_1\cdots\mu_n\alpha\beta}_{\nu_1\cdots\nu_n
\mu\nu} {S}^{\nu_1}_{\mu_1}\cdots{S}^{\nu_n}_{\mu_n}
{S_\Lambda}{}^{\mu}_{\alpha}{S_\Lambda}{}^{\nu}_{\beta}\nonumber \\
&=&-\delta^{\mu_1\cdots\mu_n\alpha\beta}_{\nu_1\cdots\nu_n \mu\nu}
\left(\bar{R}^{\nu_1}_{\mu_1}-\frac{1}{2(D-1)}\bar{R}g^{\nu_1}_{\mu_1}\right)
\cdots\left(\bar{R}^{\nu_n}_{\mu_n}-\frac{1}{2(D-1)}\bar{R}g^{\nu_n}_{\mu_n}\right)
{S_\Lambda}{}^{\mu}_{\alpha}{S_\Lambda}{}^{\nu}_{\beta}+O(h^3)\nonumber \\
&=&-(D-2)(D-3)\cdots(D-n-1)
\left(\frac{D-2}{2}\Lambda\right)^n\delta^{\alpha\beta}_{\mu\nu}
{S_\Lambda}{}^{\mu}_{\alpha}{S_\Lambda}{}^{\nu}_{\beta}+O(h^3)\nonumber \\
&=&(D-2)(D-3)\cdots(D-n-1)
\left(\frac{D-2}{2}\Lambda\right)^n L^{(2)}_{MO}{}_\Lambda+O(h^3)\nonumber \\
&=&\frac{(D-2)!}{(D-n-2)!}
\left(\frac{D-2}{2}\Lambda\right)^n \left[{\cal
R}^{\mu}_{\alpha}{\cal
R}^{\alpha}_{\mu}-\frac{D}{4(D-1)}{\cal R}^2\right]+O(h^3)
\,.
\label{3.20}
\end{eqnarray}
Thus, this is trivially of $O(h^2)$.
On the other hand, we find the equality:
\begin{eqnarray}
& &L_{MO}^{(n+2)}{}_\Lambda=
-\delta^{\mu_1\cdots\mu_n\alpha\beta}_{\nu_1\cdots\nu_n
\mu\nu} {S}^{\nu_1}_{\mu_1}\cdots{S}^{\nu_n}_{\mu_n}
{S_\Lambda}{}^{\mu}_{\alpha}{S_\Lambda}{}^{\nu}_{\beta}\nonumber \\
&=&L_{MO}^{(n+2)}-(D-n-1)(D-2)\Lambda
L_{MO}^{(n+1)}+(D-n)(D-n-1)\left(\frac{D-2}{2}\right)^2\Lambda^2
L_{MO}^{(n)}\,.
\label{eq2}
\end{eqnarray}
From (\ref{eq1}) and (\ref{eq2}), since $L_{MO}^{(1)}\propto R$ contains
at most second derivatives of $h_{\mu\nu}$ and $L_{MO}^{(2)}$ contains at
most fourth-order derivatives, we find that all $L_{MO}^{(n)}$ $(n\ge 2)$
contains at most fourth-order derivative of the metric fluctuation on the
maximally symmetric geometry.%
\footnote{If we define $L_{MO}^{(1)}=-\frac{D-2}{2(D-1)}R$ and
$L_{MO}^{(0)}=-1$, Eq.~(\ref{eq2}) holds for $n\ge 0$.}
 Moreover, because Eq.~(\ref{3.20}) shows that $L_{MO}^{(n)}{}_\Lambda$
$(n\ge 2)$ is proportional to $L_{MO}^{(2)}{}_\Lambda$, the equation of
motion can be reduced to the form similar to (\ref{2m}).

Solving the recursion relation (\ref{eq2}) with Eq.~(\ref{eq1}),
we obtain
\begin{eqnarray}
L_{MO}^{(n)}&=&\frac{(D-2)!}{(D-n)!}\left(\frac{D-2}{2}\right)^n
\left\{-D(D-1)\Lambda^n-n\Lambda^{n-1}\left[\frac{D-2}{2}\Lambda
h_{\mu\nu}h^{\mu\nu}+\frac{1}{4}h_{\mu\nu}\bar{\nabla}^2h^{\mu\nu}\right]
\right.\nonumber \\
&
&\qquad\qquad\left.+\frac{n(n-1)}{2}\Lambda^{n-2}\left[\Lambda^2h_{\mu\nu}h^{\mu\nu}
-\Lambda h_{\mu\nu}\bar{\nabla}^2
h^{\mu\nu}+\frac{1}{4}h_{\mu\nu}\bar{\nabla}^2\bar{\nabla}^2
h^{\mu\nu}\right]\right\}\,,
\end{eqnarray}
where unphysical modes are discarded, and when the background geometry, as
a solution of the equation of motion, is expressed by (\ref{bg}).

To summarize, the particle content of the the Meissner and
Olechowski gravity governed by the action
\begin{equation}
S=S_0(\Lambda')+\int d^Dx\,\sqrt{-g}\sum_{n=2}^D \alpha'_n
L_{MO}^{(n)}
\,,
\label{ss}
\end{equation}
includes two transverse traceless modes only, if $\Lambda'$ is
chosen so that the classical background solution arises as
$\bar{R}^{\mu\nu}{}_{\alpha\beta}=\Lambda
(\delta^{\mu}_{\alpha}\delta^{\nu}_{\beta}-
\delta^{\mu}_{\beta}\delta^{\nu}_{\alpha})$.%
\footnote{If we introduce an auxiliary field $s_{\mu}^{\nu}$,
$L^{(n)}_{MO}$ can be replaced by
$-\delta^{\mu_1\cdots\mu_n}_{\nu_1\cdots\nu_n}s_{\mu_1}^{\nu_1}\cdots
s_{\mu_n}^{\nu_n}-n\delta^{\mu_1\cdots\mu_{n-1}\mu}_{\nu_1\cdots\nu_{n-1}\nu}
s_{\mu_1}^{\nu_1}\cdots
s_{\mu_{n-1}}^{\nu_{n-1}}(R_\mu^\nu-\frac{1}{2(D-1)}
R\delta_\mu^\nu-s_\mu^\nu)$.}

In particular,
\begin{equation}
S_{crit}=\int d^Dx\,\sqrt{-g}\sum_{n=2}^D \alpha_n
L_{MO}^{(n)}{}_\Lambda
\,
\label{ccc}
\end{equation}
yields the critical gravity.\footnote{It is easily shown that the
background (\ref{bg}) is the solution of the equation of motion derived
from (\ref{ccc}), by doing the explicit calculation as in
Ref.~\cite{KKS}.}  Therefore this is an extension of
the critical gravity to higher order Lagrangian.
It is important to note that since the generalized Kronecker delta is
restricted by the number of dimensions, the Lagrangian (\ref{ss}) is at
most
$D$-th order in the curvatures.

\section{comment on more higher order gravity and multi-critical gravity}
The order of the Meissner and Olechowski gravity is limited by the
number of dimensions, i.e., it is valid for $n\le D$. As is obvious at a
glance,  the Lagrangian density consisting of combinations of
the type $f(R_{\mu\nu\rho\sigma})L_{MO}^{(2)}{}_\Lambda$ may lead to
the critical gravity. Note, however, that it is very complicated to find
the background metric, as the solution of the equation of motion derived
from such a general class of higher order Lagrangian.  Here, we wish to
find higher-order, symmetric invariants of curvature tensors by
systematic construction.

One should have noticed the beautiful symmetry is implemented in the
Lovelock gravity.
According to Ref.~\cite{CEP,CE,SGT2}, the Lovelock gravity on a maximally
symmetric spacetime contains a single graviton mode just as in the
Einstein gravity. Therefore we find that the Lovelock tensor
\begin{equation}
{G^{(n)}}^{\mu}_{\nu}\equiv
-\frac{1}{\sqrt{-g}}\frac{\delta(\int d^Dx\sqrt{-g}L_L^{(n)})}{\delta
g_{\rho\mu}}g_{\nu\rho}
=-2^{-(n+1)}
\delta^{\mu\sigma_1\tau_1\cdots\sigma_n\tau_n}_{\nu\lambda_1\rho_1\cdots\lambda_n\rho_n}
R^{\lambda_1\rho_1}{}_{\sigma_1\tau_1}\cdots
R^{\lambda_n\rho_n}{}_{\sigma_n\tau_n}\,
\end{equation}
is expanded in terms of the metric fluctuation $h_{\mu\nu}$ on the
background spacetime, where the Riemann tensor is given by (\ref{bg}), as
\begin{equation}
{G^{(m)}}{}^{\mu}_{\alpha}-{\bar{G}^{(m)}}{}^{\mu}_{\alpha}\propto
{\cal R}^{\mu}_{\alpha}-\frac{1}{2}{\cal R}
\delta^{\mu}_{\alpha}+O(h^2)
\,.
\end{equation}
We naturally define an extension of the Schouten tensor as
\begin{equation}
S^{\mu\nu}_{(m)}\equiv{G^{(m)}}^{\mu\nu}-\frac{1}{D-1}
G^{(m)}g^{\mu\nu}\,,
\end{equation}
where $G^{(n)}\equiv G^{(n)}{}_\rho^\rho$.
Then the following relation is obvious:
\begin{equation}
{S^{(m)}}{}^{\mu}_{\alpha}-{\bar{S}^{(m)}}{}^{\mu}_{\alpha}\propto
{\cal R}^{\mu}_{\alpha}-\frac{1}{2(D-1)}{\cal R}
\delta^{\mu}_{\alpha}+O(h^2)
\,.
\end{equation}
It is now evident by the same argument in the previous section that the
Lagrangian density consisting of a certain linear combination of the
following terms
\begin{equation}
L_{MO}^{(n\cdot m)}=-\delta^{\mu_1\cdots\mu_n}_{\nu_1\cdots\nu_n}
{S_{(m)}}^{\nu_1}_{\mu_1}\cdots{S_{(m)}}^{\nu_n}_{\mu_n}
\,
\end{equation}
leads to a model of critical gravity.%
\footnote{We should note that the auxiliary field method
\cite{NMG,NMG2,BHRT} also leads to the term which can be used in critical
higher order gravity, such as
$-s_{(n)}{}^{\mu\nu}G^{(m)}_{\mu\nu}-s_{(m)}{}^{\mu\nu}G^{(n)}_{\mu\nu}
+s_{(n)}^{\mu\nu}s_{(m)}{}_{\mu\nu}-s_{(n)}{}^{\rho}_{\rho}s_{(m)}
{}^{\rho}_{\rho}$, where $s_{(m)}^{\mu\nu}$ and $s_{(n)}^{\mu\nu}$ are
auxiliary fields.}

The gravity theory with multi-critical points, or
dubbed as `polycritical' gravity, has been proposed in
Refs.~\cite{BdHMPR,BdHMRZ,AP,Nutma,KNV}. In the theory, the analysis on
the equation of motion for the metric fluctuation reveals that there is
no scalar mode and three or more spin-two excitation due to the higher
derivatives than the sixth order.
The generalization of the construction in
Refs.~\cite{BdHMPR,BdHMRZ,AP,Nutma,KNV} to the higher order gravity can be
obtained by taking the following iterative consideration.

To obtain the equation of motion with sixth order derivatives
for the metric fluctuation,
we adopt the Lagrangian density
\begin{equation}
L_B=K^{(n)}{}^{\mu\nu}\Delta_{\mu\nu\rho\sigma}G^{(m)}{}^{\rho\sigma}\,,
\end{equation}
where 
\begin{equation}
K^{(n)}{}^{\mu\nu}\equiv-\frac{1}{\sqrt{-g}}\frac{\delta
\int d^Dx\,\sqrt{-g}\, L_{MO}^{(n)}}{\delta g_{\mu\nu}}\,,
\end{equation}
and
\begin{equation}
\Delta_{\mu\nu\rho\sigma}\equiv\frac{1}{2}
(g_{\mu\rho}g_{\nu\sigma}+g_{\mu\sigma}g_{\nu\rho})-\frac{1}{D-1}
g_{\mu\nu}g_{\rho\sigma}\,.
\end{equation}
Similarly, the eighth order differential equation for $h_{\mu\nu}$ is
derived from
\begin{equation}
L_B=K^{(n)}{}^{\mu\nu}\Delta_{\mu\nu\rho\sigma}K^{(m)}{}^{\rho\sigma}\,.
\end{equation}
To get more higher derivatives on the metric fluctuation, we repeat the
iteration
\begin{equation}
L_{N}=-\frac{1}{\sqrt{-g}}\frac{\delta
\int\,d^Dx\,\sqrt{-g}\, L_{B}}{\delta
g_{\mu\nu}}\Delta_{\mu\nu\rho\sigma}G^{(m)}{}^{\rho\sigma}\,,
\end{equation}
or
\begin{equation}
L_{NN}=-\frac{1}{\sqrt{-g}}\frac{\delta
\int\,d^Dx\,\sqrt{-g}\, L_{B}}{\delta
g_{\mu\nu}}\Delta_{\mu\nu\rho\sigma}K^{(m)}{}^{\rho\sigma}\,,
\end{equation}
and obtain the Lagrangian which yields higher order differential equation
for the metric fluctuation.
It has been explicitly confirmed that there is no scalar mode in the
theory governed by these types of the Lagrangian density  by using
conformally flat metrics  in Ref.~\cite{KKS}. 
Therefore a certain linear combinations of these terms yields polycritical
gravity, though the critical relation among the coefficients is not
pursued here.

Finally, it is interesting to point out that
these candidate Lagrangians have a form very akin to
the Lagrangian with the `detailed balance condition' of the
Ho\v{r}ava-Lifshitz gravity \cite{Horava}.

\section{Summary and conclusion}
In the present paper, we have studied a certain models with 
higher order terms in curvatures, in which excitation modes are similar to
ones in critical gravity.  We have confirmed that the Lagrangian
consisting of the Meissner-Olechowski densities yields the
higher-derivative gravity without scalar modes and have found that the
peculiar combination
$L_{MO}^{(n)}{}_\Lambda$ leads to the critical gravity with
the fourth-order derivatives on the tensor mode. We have also given a
discussion on the possible construction of terms which can be utilized in
multi-critical gravity.

In general, higher order gravity possesses many solutions.
For an interesting example, the Lagrangian consisting of a single
Meissner-Olechowski density,
$\int d^Dx\sqrt{-g}\,L_{MO}^{(n)}{}_\Lambda$, admits two
distinct solutions, in which the curvatures are
$\bar{R}^{\mu\nu}{}_{\alpha\beta}=\Lambda
(\delta^\mu_\alpha\delta^\nu_\beta-\delta^\nu_\alpha\delta^\mu_\beta)$ and
$\bar{R}^{\mu\nu}{}_{\alpha\beta}=\frac{D-2n}{D-2n-4}\Lambda
(\delta^\mu_\alpha\delta^\nu_\beta-\delta^\nu_\alpha\delta^\mu_\beta)$.
Of course, a fluctuation mode on the latter background becomes massive and
the theory on it is not critical gravity. The study of possible vacuum
transition by using the massless, massive, and log modes on the vacua
seems to be interesting, since the investigation may give a novel
cosmological evolution.

An extension of the Randall-Sundrum models \cite{RS1,RS2}, in which the
spacetime is asymptotically anti-de Sitter geometry, can be considered by
connecting the model with higher order gravity. In a general higher order
gravity, however, a thin brane require more singularity
than a usual delta function source, because of the
existence of higher derivatives on the metric fluctuation.
Only the Lovelock gravity has been applied to the Randall-Sundrum
models thus so far. A possibility of higher
order generalization is that one use two or more Meissner-Olechowski
densities for the gravitational Lagrangian, in order to cancel
the fourth-order derivative (thus this construction no longer leads to
critical gravity!). Another possibility is to consider thick
branes
\cite{LCS2,LWWZ}. Also in this case, the scalar field configuration
coupled to higher order gravity naturally leads to many vacua. Therefore,
the study of thick wall in the cosmological context is also an interesting
subject.

Because the Meissner-Olechowski gravity model does not contain the Riemann
tensors, the supersymmetrization of the model is expected to be easy at
on-shell condition in a perturbative approach. By the same reason, quantum
correction in the naive sense may also be easy to handle in the model.
Also, some compactifications and black hole solutions in the
Meissner-Olechowski gravity are interesting and worth studying in 
asymptotically anti-de Sitter spacetimes. We shall return to these various
aspects of the Meissner-Olechowski gravity in future work.

\acknowledgments
The authors would like to thank T. Zojer for appropiate information on
the work on the logarithmic modes.
We also would like to thank S. Deser and J. Edelstein for suggestion of
suitable references.




\bibliographystyle{apsrev4-1}



\end{document}